\documentclass{epl}

\usepackage{amsmath}
\usepackage{amssymb}
\usepackage{graphicx}
\usepackage{bm}
\usepackage{mathrsfs}

\newcommand{\msl}{\mathscr{L}}
\newcommand{\msf}{\mathscr{F}}
\newcommand{\msg}{\mathscr{G}}

\title{The Intense Radiation Gas}
\shorttitle{The intense radiation gas}

\author{M.\ Marklund\inst{1,2,3} \and P.K.\ Shukla\inst{1,2,3} 
  \and B. Eliasson\inst{3}}

\institute{
  \inst{1} Department of Physics, Ume{\aa} University, 
    SE--901 87 Ume{\aa}, Sweden \\
  \inst{2} Centre for Fundamental Physics,
  Rutherford Appleton Laboratory,
  Chilton Didcot, Oxfordshire, OX11 0QX, UK \\
  \inst{3} Institut f\"ur Theoretische Physik IV, Fakult\"at f\"ur
    Physik und Astronomie,  Ruhr-Universit\"at Bochum, D--44780 Bochum,
    Germany
} 


\begin{document}

\maketitle 

\date{\today}

\begin{abstract}
We present a new dispersion relation for photons that are nonlinearly 
interacting with a radiation gas of arbitrary intensity due to photon--photon 
scattering. It is found that the photon phase velocity decreases with
increasing  
radiation intensity, it and attains a minimum value in the limit of
super-intense  
fields.  By using Hamilton's ray equations, a self-consistent kinetic
theory for  
interacting photons is formulated. The interaction between an electromagnetic pulse 
and the radiation gas is shown to produce pulse self-compression and nonlinear 
saturation. Implications of our new results are discussed.       
\end{abstract}

In classical electrodynamics photons are indifferent to each other in vacuum, 
while in quantum electrodynamics (QED) photons can interact via virtual 
electron--positron pairs, giving rise to vacuum photon--photon scattering
\cite{Heisenberg-Euler}. This is commonly expressed using a series expansion 
(in the field strength) of the Heisenberg--Euler Lagrangian, yielding
nonlinear corrections to Maxwell's vacuum equations.
These corrections give rise to single particle effects (some of which are highly speculative), 
such as 
closed photon paths \cite{Novello-etal}, vacuum birefringence 
\cite{Bialynicka-Birula,Adler,Heyl-Hernquist}, photon splitting \cite{Adler} and lensing effects 
in a strong magnetic field \cite{Harding,DeLorenci-etal}, 
and coherent effects, such as the self-focusing 
of beams \cite{Soljacic-Segev} or the formation of light bullets
\cite{Brodin-etal}. Moreover, as an example of a collective effect,
the QED photon self-energy will change the refractive index of a radiation 
gas \cite{Barton,Gies,Thoma}, which can result in, e.g., Cherenkov emissions in a
radiation gas \cite{NJP}.  

Recently, it has been shown  that nonlinear vacuum corrections 
cause photonic collapse in a radiation gas \cite{Marklund-Brodin-Stenflo}. 
The phenomena of photonic collapse in two space-dimension has been
confirmed by computer simulation studies \cite{Shukla-Eliasson}. 
As indicated in Ref.\ \cite{Shukla-Eliasson}, the increase in intensity 
will show no upper bound as the collapse has begun, and will
eventually surpass the Schwinger field $\sim 10^{16}\,\mathrm{V/cm^2}$. 
Even as dispersive corrections are added
\cite{Rozanov,Marklund-Eliasson-Shukla,Shukla-etal},  
with resulting pulse splitting, the field intensities within the
collapse region may  still reach values well above the ones allowed by the 
approximate Heisenberg--Euler formalism. Thus, this raises the
critical question   
of how far the predictions of the approximate Heisenberg--Euler Lagrangian 
can be trusted, i.e.  whether the collapse of electromagnetic fields is 
a ``true'' effect or an artifact of the approximation used. 

The investigation of the occurrence of singularities is important not 
only because of its theoretical interest, but also by the fact that Nature 
may offer the environment for such events to take place. Within astrophysical 
and cosmological settings, such as neutron stars and magnetars
\cite{Kouveliotou},  
the conditions are met for the ensue of collapse of disturbances in the
photon distribution. These conditions may also soon be met in
laboratory environments. Today's lasers can produce $10^{21}-10^{22}$
W/cm$^2$ \cite{Mourou-Barty-Perry}, and the next generation laser-plasma
systems are expected to reach $10^{25}$ W/cm$^2$.
\cite{cai04,bob,bob2}. This has been used to suggest that field
strengths close to the Schwinger critical value could be reached
\cite{Bulanov-Esirkepov-Tajima}. Thus, here the nonlinear effects
introduced by photon--photon scattering will be significant, and the
weak field approximation no longer holds. 
 
In this Letter, we investigate a photon gas of arbitrary intensity 
at the one-loop level. We begin with a discussion of the dispersion relation 
for photons travelling in a crossed-field configuration, where after we 
derive a new expression for the phase velocity of photons in a  
radiation gas. In the weak field limit, our new expression exactly reproduce 
the previously known results for the phase velocity, while the limit of an 
ultra-intense radiation gas gives rise to a constant subluminal phase
velocity.  
It is argued that superluminal velocities are prohibited by pair creation. Our 
general dispersion relation shows that the effective force between the
photons will always be attractive, similar to the weak field case. 
Thus, the collapse of inhomogeneities can occur within a perturbed
gas of intense photons. Moreover, we investigate the implication for
the collective dynamics of photons, using a kinetic theory. A set of equations
governing the dynamics of an interacting gas and an electromagnetic
pulse is derived. It is shown that they admit pulse self-compression and
nonlinear saturation. The significance of our results are discussed, together 
with some possible applications. 

The effective field theory of soft photon--photon scattering in constant 
background electromagnetic fields can be formulated using the Lagrangian 
density $\msl = \msl_0 + \msl_{c}$, where $\msl_0 = -\mathscr{F}$ is the 
classical free field Lagrangian, and the one-loop correction is
\cite{Schwinger}   
\begin{eqnarray}
  \msl_{c} = -\frac{1}{8\pi^2}\int_0^{\mathrm{i}\infty}
  \frac{ds}{s^3}\mathrm{e}^{-m_e^2s} \Big[
  (es)^2ab\,\coth(eas)\,\cot(ebs)
  - \tfrac{1}{3}(es)^2(a^2 - b^2) - 1 \Big] ,
\label{eq:lagrangian}
\end{eqnarray}
where $m_e$ is the electron mass, $e = |e|$ is the electron charge, $a = 
[(\msf^2 + \msg^2)^{1/2} + \msf]^{1/2}$, $b = [(\msf^2 +
\msg^2)^{1/2} - \msf]^{1/2}$, $\msf \equiv F_{ab}F^{ab}/4 = 
(\bm{B}^2 - \bm{E}^2)/2$, $\msg \equiv
F_{ab}\widehat{F}^{ab}/4 = -\bm{E}\cdot\bm{B}$, 
$\widehat{F}^{ab} = \epsilon^{abcd}F_{cd}/2$. Thus, $\msf = (a^2 -
b^2)/2$ and $|\msg| = ab$. 

The dispersion relation for a test photon in a background
electromagnetic field is given by \cite{Bialynicka-Birula} (see also Ref.\ \cite{Dittrich-Gies2}) 
\begin{eqnarray}
  \left[1 + \tfrac{1}{2}{\lambda}(\bm{E}^2 + \bm{B}^2)\right]v^2 -
  2{\lambda}\hat{\bm{k}}\cdot(\bm{E}\times\bm{B})v
  + {\lambda}\left[ \tfrac{1}{2}(\bm{E}^2 + \bm{B}^2) -
  (\hat{\bm{k}}\cdot\bm{E})^2 - 
  (\hat{\bm{k}}\cdot\bm{B})^2\right]  - 1 = 0,
\label{eq:disprel}
\end{eqnarray}
where $v = \omega/|\bm{k}| \equiv 1/n$ is the normalized
(by the speed of light) photon phase velocity, 
$n$ is the refractive index, $\omega$ is the photon
frequency, $\bm{k}$ is the photon wavevector, and $\hat{\bm{k}} =
\bm{k}/|\bm{k}|$. The effective action charge ${\lambda}$ is \cite{Bialynicka-Birula} (see also 
\cite{Dittrich-Gies2}) 
\begin{equation}
  {\lambda} = \frac{(\partial^2_{\msf} +
  \partial^2_{\msg})\msl}{-2\partial_{\msf}\msl +
  \msf(\partial^2_{\msf} + \partial^2_{\msg})\msl -
  2(\msf\partial^2_{\msf} + \msg\partial^2_{\msf\msg})\msl} .
\label{eq:Q}
\end{equation}

To start with, we look at crossed fields for which  
$|\bm{E}| = |\bm{B}|$ and $\bm{E}\cdot\bm{B} = 0$, i.e. $a = b = 0$. 
We define $\hat{\bm{s}} = (\bm{E}\times\bm{B})/|\bm{E}|^2$, and assume 
that the test photon is a part of an ensemble of random photons such 
that $\langle \hat{\bm{k}}\cdot\hat{\bm{s}}\rangle = 0$ and $\langle
(\hat{\bm{k}}\cdot\hat{\bm{s}})^2\rangle = 1/3$, where the angular
bracket denotes an ensemble average. Then, Eq.\ (\ref{eq:disprel})
yields the average phase velocity 
$  v = [(1 - \tfrac{1}{3}{\lambda}_{\mathrm{cf}}|\bm{E}|^2)/
(1 + {\lambda}_{\mathrm{cf}}|\bm{E}|^2)]^{1/2}$,   where the index 
$cf$ denotes the crossed-field value.  Taking the limits  $a \rightarrow b$ 
and $b \rightarrow 0$ in Eq.\ (\ref{eq:Q}) and using the Lagrangian 
(\ref{eq:lagrangian}), we obtain 
\begin{equation}
  {\lambda}_{\mathrm{cf}}^{-1} 
  =
  \frac{45}{22}\frac{4\pi}{\alpha}\mathscr{E}_{\text{crit}} ,  
\label{eq:lambda}
\end{equation}  
corresponding to field strengths $\sim
8\times10^{17}\,\,\mathrm{V/cm}$ or $3\times10^{15}$ G. Here,
$\mathscr{E}_{\text{crit}} = E_{\text{crit}}^2$, where $E_{\text{crit}} 
\equiv m_e^2/e =  m_e^2c^3/e\hbar \sim 10^{16}\,\, \mathrm{V/cm}$
denotes the critical Schwinger field and $\alpha = e^2/4\pi$ is the
fine-structure constant. Note that this is the same charge as the 
geometrical average of the coefficient obtained from the polarisation 
tensor in the weak field limit \cite{Bialynicka-Birula}, 
i.e. an average over polarisation states. Moreover, 
the Schwinger critical field will, in our case, not correspond
to a threshold for the pair-creation. 

We will now analyse the radiation gas.   
An electromagnetic wave moving in an isotropic and homogeneous medium
with the refractive index $1/v$ will satisfy $|\bm{B}| = |\bm{E}|/v$, 
$\msg = - \bm{E}\cdot\bm{B}$ = 0, and 
$\msf = \tfrac{1}{2}(v^{-2} - 1)|\bm{E}|^2 \geq 0$; the latter
inequality holding for a subluminal phase velocity. This implies $a
\neq 0$ and $b = 0$. For superluminal velocities, the invariants satisfy 
$a = 0$ and $b \neq 0$, for which 
the spontaneous pair production is predicted \cite{Schwinger}. 
We will assume $v \leq 1$
throughout. A gas of photons may be viewed as an ensemble $\{\bm{E}_i,
\bm{B}_i \}$ interacting, due to elastic photon--photon scattering,
through the refractive index of the gas. 
The effective action charge, as given by the
expression (\ref{eq:Q}), will now depend on both the field strength
and the phase velocity of the gas. But through (\ref{eq:disprel}) the 
phase velocity itself depends on ${\lambda}$, and is 
therefore implicitly determined via (\ref{eq:disprel}). 
We assume that the momentum flux in the gas rest frame vanishes, i.e.
$\langle\bm{E}\times\bm{B}\rangle = 0$, and the gas is characterized
by its energy density $\mathscr{E} \equiv
\langle|\bm{E}|^2\rangle$. We, therefore, have a nonzero invariant
according to $a = [(v^{-2} - 1) \mathscr{E} ]^{1/2}$.
 
\begin{figure}
  \includegraphics[width=.8\columnwidth]{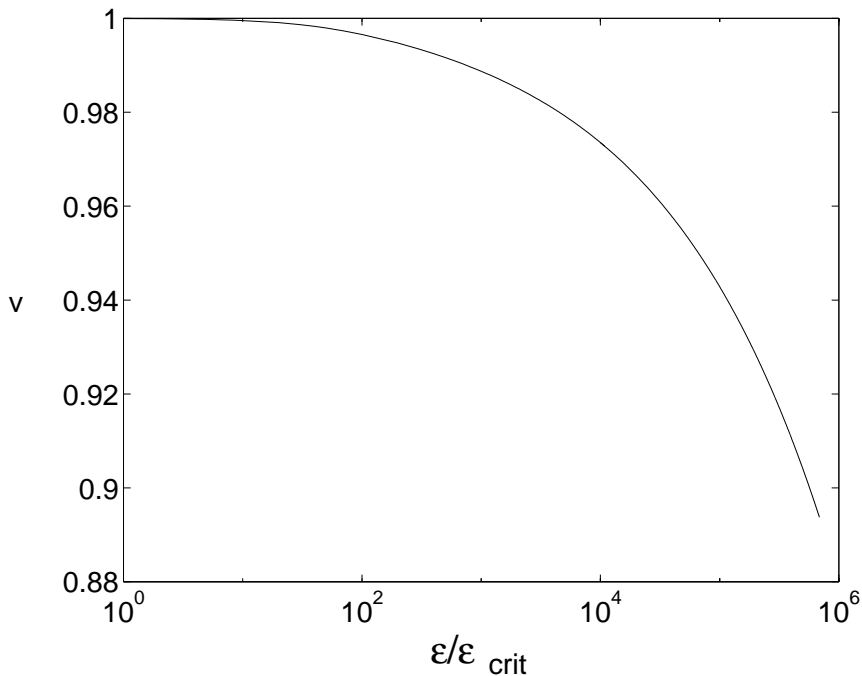}
  \caption{The normalized phase velocity $v$, as given by Eqs.\
    (\ref{eq:refractive})--(\ref{eq:Qgas}), plotted as a function of the
    normalised energy density $\mathscr{E}/\mathscr{E}_{\text{crit}}$.}
\label{fig:velocity}
\end{figure}

Equation (\ref{eq:disprel}) may now be solved to give the expression  
\begin{equation}
  v^2 = \frac{1 - \tfrac{2}{3}{\lambda}_{\text{gas}} \mathscr{E} + \sqrt{1 -
  2{\lambda}_{\text{gas}}\mathscr{E} +
  \tfrac{1}{9}({\lambda}_{\text{gas}}\mathscr{E})^2}}{2 + 
  {\lambda}_{\text{gas}} \mathscr{E}} 
\label{eq:refractive}
\end{equation}
for the phase velocity $v$. 
An expression of the same functional form can be derived for a random
set of photons in the field of a plane wave pulse with $\mathscr{E}$
replaced by $|E_p|^2$, where $E_p$ being the slowly varying pulse envelope. 
The field space Laplacian operating on the Lagrangian density $\msl$
yields (using $z \equiv a e s$)
\begin{equation}
  \lim_{b\rightarrow 0}\,(\partial^2_{\msf} + \partial^2_{\msg})\msl =
  \frac{1}{8\pi^2}\frac{e^2}{2 a^2}
  \int_0^{\mathrm{i}\infty} \frac{dz}{z}\,\mathrm{e}^{-E_{\text{crit}}
  z/a} \left( \frac{1 - z\coth z}{\sinh^2 z} + \frac{1}{3}z\coth z 
  \right) \equiv \frac{\alpha}{4\pi}\frac{1}{a^2}F(a/E_{\text{crit}}), 
\label{eq:FG}
\end{equation}
and gives the behavior of the effective action charge (\ref{eq:Q}),
while $\lim_{b\rightarrow 0}\,[ -2\partial_{\msf}\msl_c
  -a^2\partial^2_{\msf}\msl_c ] \equiv 
  (\alpha/8\pi)G(a/E_{\text{crit}})$ gives a small correction, and can
safely be neglected. In its full form, Eq.\ (\ref{eq:Q}) can be
written  as
\begin{equation}
  {\lambda}_{\text{gas}} = \frac{\alpha}{4\pi
  a^2}\,\frac{F(a/E_{\text{crit}})}{2 +  
  (\alpha/8\pi)[ F(a/E_{\text{crit}}) +
  G(a/E_{\text{crit}}) ]} . 
  \label{eq:Qgas}
\end{equation}
The same functional relationship is found for a random set of photons in the
field of a plane wave pulse, with $\mathscr{E} \rightarrow |E_p|^2$. 
The full normalized phase velocity $v$, as given by Eqs.\
(\ref{eq:refractive})--(\ref{eq:Qgas}), is plotted in Fig.\
\ref{fig:velocity} as a function of the normalised energy
$\mathscr{E}/\mathscr{E}_{\text{crit}}$. We see that the normalized phase
velocity goes to unity for small energy densities, and decreases with
increasing energy densities.  
In the weak field limit ($ \mathscr{E} / \mathscr{E}_{\text{crit}}
\ll 1$), we take $v \simeq 1$ in the evaluation of
${\lambda}_{\text{gas}}$,  
yielding ${\lambda}_{\text{gas}} \simeq {\lambda}_{\mathrm{cf}}$ 
[see Eq.\ (\ref{eq:lambda})]. A series expansion of Eq.\ 
(\ref{eq:refractive}) yields   
$ v \simeq 1 - \tfrac{2}{3}{\lambda}_{\mathrm{cf}} \mathscr{E}$,  
consistent with the results in
Refs.\
\cite{Marklund-Brodin-Stenflo,Dittrich-Gies2,Bialynicka-Birula}. On
the other hand, in the ultra-strong field limit 
($ \mathscr{E} / \mathscr{E}_{\text{crit}} 
\gg 1$), $F$ goes to infinity in Eq.\ 
(\ref{eq:Qgas}), giving ${\lambda}_{\text{gas}} \simeq 2a^{-2}
= 2v^2[(1  - v^2) \mathscr{E} ]^{-1}$. Using Eq.\
(\ref{eq:refractive}), we obtain the asymptotic phase velocity 
\begin{equation}
v_{\infty} \simeq 1/\sqrt{5} \approx 0.45 .
\label{eq:constant}
\end{equation}
We stress that, formally, the one-loop approximation can be viewed as 
questionable in the limit $\mathscr{E} \rightarrow \infty$. However, it is 
straightforward to show, using higher order perturbation theory \cite{Ritus}, that the 
renormalization group improved Lagrangian will include at most a logarithmic 
diverging behaviour, and thus the one-loop approximation gives a very
accurate description even in the limit highly super-critical intensities
(as in the case of pure magnetic fields \cite{Dittrich-Gies})

Under most circumstances, the contribution proportional to $\alpha$ in 
the denominator of Eq.\ (\ref{eq:Qgas}) is small and may be
neglected. When $\mathscr{E}/\mathscr{E}_{\text{crit}} \geq
1$, we obtain $F \simeq a/3E_{\text{crit}}$ from Eq.\ (\ref{eq:FG}), 
and we have  
\begin{equation}
  {\lambda}_{\text{gas}} 
  \simeq \frac{\alpha}{8\pi a^2}F(a/E_{\text{crit}}) 
  \simeq \frac{\alpha}{24\pi E_{\text{crit}}} \frac{v}{\sqrt{1 -
  v^2}} \frac{1}{\sqrt{\mathscr{E}}} \ .
\label{eq:approxlambda}
\end{equation} 

Our knowledge of the nonlinear refractive properties of the radiation
gas gives a means for investigating the effects of higher-order
nonlinear corrections to the standard first order Heisenberg--Euler
Lagrangian, and probe the significance of higher-order effects for
photonic collapse 
\cite{Marklund-Brodin-Stenflo,Shukla-Eliasson,Marklund-Eliasson-Shukla}. 
The dynamics of coherent photons, travelling through an intense
radiation gas, may be analysed following Ref.\
\cite{Marklund-Brodin-Stenflo}. We obtain a
nonlinear Schr\"odinger equation for the slowly varying
pulse envelope $E_p$ according to \cite{Kivshar-Agrawal}   
\begin{equation}
  i\left(\frac{\partial}{\partial t} + \bm{v}_0\cdot\bm{\nabla}
  \right)E_p + \frac{v_0}{2k_0}\nabla_{\perp}^2 E_p + 
  \omega_0 \frac{n_{\mathrm{nl}}(\delta\mathscr{E})}{n_0} E_p = 0 ,
\label{eq:nlse}
\end{equation}
where the subscript $0$ denotes the background state, 
$\nabla_{\perp}^2 = \nabla^2 -(\hat{\bm{k}}_0\cdot\bm{\nabla})^2$, 
$\delta\mathscr{E} = \mathscr{E} - \mathscr{E}_0$ is a
perturbation,  
$v_0 = v(\mathscr{E}_0)$, $n_0 = n(\mathscr{E}_0)$, 
$n_{\mathrm{nl}}(\delta\mathscr{E}) = \sum_{m = 1}^{\infty}
n_0^{(m)}\delta\mathscr{E}^m/ m!$, and $n_0^{(m)} = d^m n_0/d
\mathscr{E}_0^m$. 

For a dispersion relation $\omega = |\bm{k}|v(\bm{r}, t)$, the motion 
of a single photon may be described by the Hamiltonian ray equations \cite{Mendonca}
\begin{equation}
  \dot{\bm{r}} = \partial_{\bm{k}}\omega =
  v\,\hat{\bm{k}}, \,\,  \text{ and }\,\,  
  \dot{\bm{k}} = -\bm{\nabla}\omega =
   \omega \bm{\nabla} \ln n .
\label{eq:hamilton}
\end{equation}
Since $n = n(\mathscr{E})$ and $d n/d \mathscr{E} > 0$ always holds, a denser 
region of the radiation gas will exercise an attractive force on the photon
\cite{Partovi}, thus creating lensing effects. The single particle dynamics 
thus supports that photonic self-compression is an inherent property
of the one-loop radiation gas, but we note that as the density of a region
increases, the phase velocity approaches a constant value (\ref{eq:constant}), 
i.e. $\bm{\nabla} \ln n \rightarrow 0$. 

Using Eq.\ (\ref{eq:hamilton}), the collective interaction of photons
can then be formulated as~\cite{Mendonca}
\begin{equation}\label{eq:kinetic}
  \frac{\partial N_k}{\partial t} + v\hat{\bm{k}}\cdot\bm{\nabla}N_k 
  + \frac{\omega}{n}(\bm{\nabla}n)\cdot
  \frac{\partial 
  N_k}{\partial\bm{k}} = 0 , 
\end{equation}
where the distribution function $N_k(\bm{r}, t)$ has been
normalised such that  $\int N_k(\bm{r},t)\,d\bm{k}$ is the number
density. Interestingly enough, in the limit of large intensity 
$\mathscr{E} \gg \mathscr{E}_{\text{crit}}$, since the phase velocity
approaches a constant value, the one-loop radiation gas
essentially evolves by free streaming, making higher-order
loop corrections important in this extreme regime \footnote{%
  The two-loop corrections are normally very small compared
  to the one-loop Lagrangian \cite{Ritus}, making higher-order loop
  corrections important in the regime where $v$ is close to the
  constant value (\ref{eq:constant})}.   

Following Ref.\ \cite{Marklund-Brodin-Stenflo}, the response of the
radiation gas to a plane wave pulse may be formulated in terms of an
acoustic wave equation 
\begin{equation}
  \left(  \frac{\partial^2}{\partial t^2} - \frac{v_0^2}{3}\nabla^2
  \right)\delta\mathscr{E} = - \frac{\mathscr{E}_0}{n_0} \left(
  \frac{\partial^2}{\partial t^2} + v_0^2 \nabla^2
  \right)n_{\mathrm{nl}}(|E_p|^2 ) ,    
\label{eq:acoustic}
\end{equation}
derived from Eq.\ (\ref{eq:kinetic}) by taking the moments, using the
equation of state $P = \mathscr{E}/3$ ($P$ being the pressure), and
linearizing around the background state. If the time response of
$\delta\mathscr{E}$ is slow, Eq.\ (\ref{eq:acoustic}) gives
\begin{equation}
  \delta\mathscr{E} 
  \simeq \frac{3\mathscr{E}_0n_0'}{n_0} \left( 
   1 + \frac{n_0''}{2 n_0'}|E_p|^2 \right) |E_p|^2 .
\label{eq:deltaE}
\end{equation}
Using (\ref{eq:deltaE}) and the expression for
$n_{\mathrm{nl}}(\delta\mathscr{E})$, we can write Eq.\ (\ref{eq:nlse}) as 
\begin{eqnarray}
  i\left( \frac{\partial}{\partial t} + \bm{v}_0\cdot\bm{\nabla}
  \right)E_p + 
  \frac{v_0}{2k_0}\nabla_{\perp}^2 E_p 
  +
  \omega_0\left(\frac{3\mathscr{E}_0 n_0'}{n_0}\right)^2\!\!\!
  \left(
   1 + \frac{n_0''}{2 n_0'}|E_p|^2\right)|E_p|^2 E_p = 0 .
\label{eq:nlse2}
\end{eqnarray}
When ${n_0''}|E_p|^2/{2 n_0'} \ll 1$, we have a self-focusing
nonlinearity in Eq.\ (\ref{eq:nlse2}), but as $|E_p|$ grows the
character of the nonlinear coefficient changes. The coefficient is
positive when $|E_p|^2 < E_S^2 \equiv |2n_0'/n_0''|$, but since
$n_0'' < 0$ for all 
$\mathscr{E}_0$, the sign will change as the pulse amplitude grows
above the saturation field strength $E_S$, making the nonlinearity
defocusing and arresting the collapse. 
The numerical value of this turning point is
dependent on the background parameter $\mathscr{E}_0$. For low
intensity radiation gases, $n_0'' \simeq 0$, and Eq.\
(\ref{eq:nlse2}) always displays self-focusing, i.e. the field
strengths can reach values above the Schwinger field. When
$\mathscr{E}_0$ roughly reaches the critical value
$\mathscr{E}_{\text{crit}}$, the weak field 
approximation breaks down, and Eqs.\
(\ref{eq:refractive}) and (\ref{eq:approxlambda}) can be used to
derive an expression for $E_S$. 
As an example displaying the general character of the intense
background case, consider $\mathscr{E}_0 = \mathscr{E}_{\text{crit}}\times
10^{2}$. We find that $E_S \simeq 2\times10^{17}\,\mathrm{V/cm}\, >
E_{\text{crit}}$, i.e. the pulse saturates above the Schwinger
critical field. Thus, both the weak and moderately strong intensity
cases, as described by Eq.\ (\ref{eq:nlse2}), display 
self-compression above the Schwinger critical field. 
The analysis of the saturation field strength for arbitrary
background intensities requires that we 
abandon the approximation (\ref{eq:approxlambda}), and take the full  
denominator of Eq.\ (\ref{eq:Qgas}) into account. 

An astrophysical setting where photon--photon 
scattering could result in photonic self-compression is the magnetar 
\cite{Kouveliotou}. It has been shown that the quakes from crust tension, 
due to the extreme fields $\sim 10^{15}\,\mathrm{G}$, would release large 
amounts of low-frequency photons, creating a high intensity photon gas
\cite{Kondratyev}, suitably described within the framework developed
here. The full consideration of such environments is beyond the scope
of the present paper, and is left for future research.

At the laboratory level, laser intensities are approaching
$10^{22}\,\, \mathrm{W/cm^{2}}$ and beyond 
\cite{Mourou-Barty-Perry}. While there are upper 
limits to powers achievable by ordinary laser techniques
\cite{Mourou-Barty-Perry}, laser-plasma
systems hold the promise of surpassing these limits, coming closer to
the Schwinger intensity $10^{29}\,\, \mathrm{W/cm^{2}}$
\cite{Bulanov-Esirkepov-Tajima}. Due to the relativistic ponderomotive
force, the plasma particles will be expelled from the high
intensity regions \cite{Yu-etal}, and plasma channels will be
created. As shown in Ref.\ \cite{Shen-Yu}, these channels provide the
right conditions for photon--photon scattering to take
place, and it has been suggested to use
this a means for detecting elastic scattering between photons
\cite{Shen-etal}. In the next generation laser-plasma systems, the
radiation intensity within the plasma channels will reach levels where
radiation vacuum collapse can take place. In such systems, 
the full nonlinear structure of the vacuum, as given by its
refractive properties [Eq.\ (\ref{eq:refractive})], will be
important. Moreover, it has been shown that once the pulse satisfies a
collapse criteria, a self-compression of radiation gases will
naturally occur due to QED effects
\cite{Marklund-Brodin-Stenflo,Shukla-Eliasson,%
Marklund-Eliasson-Shukla,Shukla-etal},   
making it possible to reach, within the self-consistent theory of
elastic photon--photon scattering, the necessary energy scales for the
novel  strong-field effects to take place. As we have seen, the weak field
approximation gives a good description up to $\mathscr{E}_0 \sim
\mathscr{E}_{\text{crit}}$, thus allowing self-compression of a plane
wave pulse above the critical field strength $E_{\text{crit}}$. On the
other hand, when $\mathscr{E}_0 > \mathscr{E}_{\text{crit}}$, the pulse will
be saturated by higher order nonlinearities, but still above the
critical field strength $E_{\text{crit}}$. Thus, there is hope that
one could probe the salient features of the nonlinear quantum vacuum
in the near future.\footnote{It should be mentioned that derivative correction
due to rapidly varying fields could be of importance in certain applications
\cite{Tsai-Erber},
e.g.\ involving $\gamma$-rays, but has little effect in the cases discussed here.} 

To summarize, we have presented a new dispersion relation for a test 
photon in a radiation gas background, accurate to the one-loop level and 
valid for {\it an arbitrary radiation gas intensity}. We numerically analyzed 
the properties of the refractive index, and demonstrated photonic lensing 
due to energy density inhomogeneities. Furthermore, using the Hamiltonian ray 
equations we developed a kinetic theory for photon-photon interactions, and 
found that the full nonlinear one-loop theory supports the notion of photonic 
self-compression up to a characteristic energy density where higher-order 
vacuum nonlinearities yield saturation. Moreover, the value of the saturated field 
strength was shown to depend on the radiation gas background energy density,
and that it could surpass the Schwinger critical field strength.    

\acknowledgments
This work was supported by the European Commission (Brussels, Belgium)
through Contract No.\ HPRN-2001-0314. M. M.\ would like to
thank Chris Clarkson, Gert Brodin, Lennart Stenflo and
Pontus Johannisson for helpful discussions.

\end{document}